\documentclass[12pt]{iopart}
\usepackage[dvips]{graphicx}

\begin{document}
\title{Charging Induced Emission of Neutral Atoms from NaCl Nanocube Corners}
\author{Davide Ceresoli$^{1,\dag}$ and Tatyana Zykova-Timan$^{1,\ddag}$}
\address{$^1$ Scuola Internazionale Superiore di Studi Avanzati (SISSA)
and DEMOCRITOS, via Beirut 2, 34014 Trieste, Italy}
\address{$\dag$ Present address: Dept. of Materials Science and
Engineering, Massachusetts Institute of Technology (MIT), 77 Massachusetts
Avenue, Cambridge MA 02139-4307, USA}
\address{$\ddag$ Present address: Institut f\"ur Physik, Johannes
Gutenberg-Universit\"at Mainz, Staudingweg 7, 55099 Mainz, Germany}
\author{Erio Tosatti$^{1,2}$}
\address{$^1$ Scuola Internazionale Superiore di Studi Avanzati (SISSA)
and DEMOCRITOS, via Beirut 2, 34014 Trieste, Italy}
\address{$^2$ International Center for Theoretical Physics (ICTP),
Strada Costiera 11, 34014, Trieste, Italy}\ead{tosatti@sissa.it}

\begin{abstract}
Detachment of neutral cations/anions from solid alkali halides can
in principle be provoked by donating/subtracting electrons to the
surface of alkali halide crystals, but generally constitutes a very
endothermic process. However, the amount of energy required for
emission is smaller for atoms located in less favorable positions,
such as surface steps and kinks. For a corner ion in an alkali halide
cube the binding is the weakest, so it should be easier to remove that
atom, once it is neutralized. We carried out first principles density
functional calculations and simulations of neutral and charged NaCl
nanocubes, to establish the energetics of extraction of neutralized
corner ions. Following hole donation (electron removal) we find that
detachment of neutral Cl corner atoms will require a limited energy of
about 0.8~eV. Conversely, following the donation of an excess electron
to the cube, a neutral Na atom is extractable from the corner at the
lower cost of about 0.6~eV. Since the cube electron affinity level
(close to that a NaCl(100) surface state, which we also determine) is
estimated to lie about 1.8~eV below vacuum, the overall energy balance
upon donation to the nanocube of a zero energy electron from vacuum will
be exothermic. The atomic and electronic structure of the NaCl(100)
surface, and of the nanocube Na and Cl corner vacancies are obtained
and analyzed as a byproduct.
\end{abstract}
\pacs{73.22-f, 71.15-m, 61.46.Bc, 68.43.Rs}
\maketitle

\section{Introduction}\label{sec:intro}
Displacement, or detachment of ions, atoms or molecules from a bulk
alkali halide crystal is generally an energetically costly process. For
instance, the energy required to remove a NaCl molecule from bulk NaCl is
$\sim$1.8~eV per molecule~\cite{CRC}. Nonetheless, following addition or
removal of an electron to the crystal, individual neutral atoms may be
extracted from an alkali halide surface with greater ease. On account
of the absence of Coulomb binding, and of the resulting low stability
of the surface neutralized atom, such a process will be energetically
less expensive.

For the extraction of a neutral halogen atom there is the initial cost of
the preliminary neutralization process, requiring roughly the exciton
energy of NaCl of order 8~eV (Ref.~\cite{zielasek00}). Once primed
with this (surface) exciton the alkali halide solid is known to emit
a halogen atom with creation of a surface F-center~\cite{georgiew88,
li92,puchin93,puchin94,rohlfing05}.  Recent experiments with STM tips
on NaCl(100) atomically thin films~\cite{meyer05} showed in addition
that it is feasible to extract halogen atoms under a tip without prior
excitation but in conditions of applied voltage.

For the extraction of neutral Na atoms the preliminary neutralization
process is on the contrary exoenergetic, since an electron initially
at the energy of vacuum zero can lower its energy by falling in the
affinity level of the alkali halide surface, or of a defect, as will be
discussed below.

Here we examine theoretically the emission of both neutral Cl and
Na atoms upon subtraction or donation respectively of electrons to
NaCl cubic nanoclusters.  Experimentally, neutral NaCl nanoclusters
may acquire negative charge through electron attachment, or positive
charge via electron photoemission~\cite{conover88}.  Early empirical
estimates suggested the possible thermal emission of neutral Na
atoms from NaCl nanoclusters with an activation energy as small as
0.4~eV. An even smaller energy of about 0.04~eV was estimated for the
detachment of neutral Cl atoms~\cite{martin83,diefenbach85,galli86}.
Such low detachment energies are likeliest for the weakly bound atoms
at the cluster corners. Interestingly, since as mentioned above electron
attachment to the NaCl affinity level will itself liberate energy, this
suggests the possibility to generate cheaply reactive alkali atoms whose
availability might be of considerable importance for principle as well
as for practical applications.

In this paper we use first principles density functional theory (DFT)
methods to calculate the energy necessary to detach a neutral corner
alkali (halogen) atom after donation (subtraction) of an electron to an
alkali halide crystalline nanocube. The calculated neutral Na extraction
energy is 0.6--0.7~eV, somewhat larger than, but of similar magnitude
of, the semi-empirical estimates. The extraction energy obtained for
neutral Cl is about 0.8~eV, much larger than the  empirical estimates.
In the process, we independently clarify several aspects of NaCl(100)
surface electronic structure.

Carrying initially out our calculations for an infinite flat, fully
relaxed NaCl(100) surface, we first of all identify the surface
electronic states, focusing on those that lie energetically inside the
bulk energy gap. To a zero order approximation, these states foreshadow
the wavefunctions which an added hole or an extra electron would occupy in
a hypothetical fully delocalized, plane-wave like state when added to the
infinite, perfect NaCl(100) surface. Thus these filled and empty surface
qualitatively describe respectively the ideal ionization and affinity
levels of the infinite NaCl solid, when bounded by NaCl(100) surfaces.

Actually however, a hole/electron initially created in such a surface
state may not generally remain stable in such a plane-wave like,
delocalized state.  Even in the absence of defects and of extrinsic
traps, electron-phonon coupling alone can lead to a spontaneous local
deformation of the surface lattice structure, whereby the surface electron
or hole may become localized, or ``self-trapped''.  A self-trapped hole
near a Cl$^-$ surface ion, or a self-trapped electron near a Na$^+$
surface ion can lead to ion neutralization, canceling the electrostatic
attraction which originally tied that ion to surface lattice. The surface
atom thus neutralized will in that case remain bound to the surface only
through weak induced-dipole and dispersion forces, now relatively easy to
break. We will not actually pursue the existence of such ideal surface
self-trapped states here. The reason for this is that in presence of
such omnipresent surface defects as steps, kinks, or corners, the actual
trapping of electrons and holes will take place right there, with great
preference over possible self-trapping in the flat surface regions.

As a preliminary step, the physics and the energetics of electron/hole
capture will be addressed in this paper by direct electronic structure
calculations of initially neutral, then charged, NaCl(100) nanocubes. The
cubes are chosen with ``magic structures''~\cite{martin83} of increasing
size, namely (NaCl)$_4$ (NC4) and (NaCl)$_{32}$ (NC32). Following the
charging of the nanocube, yielding NC4$^\pm$ and NC32$^\pm$, one corner
atom, Cl in the positive case, Na in the negative case, is gradually
and adiabatically forced to detach, by moving along a (111) direction
away from the cube corner. The detaching Cl and Na atoms have captured
the added hole or electron and are spontaneously neutral, while the
cluster left behind has an ion vacancy and is charged. The increase of
total energy upon detachment defines the energy barrier for neutral atom
escape. We find that the results for the two cluster sizes, NC4 and NC32
essentially coincide, suggesting that size effects are negligible, so that
the resulting calculated detachment energies can be trusted as a good
approximation the cost of neutral corner atom extraction for arbitrary
sizes, and probably even from corners of more general geometries.

These results now permit a quantitative discussion of the overall
energy balance of the neutral atom emission process and in particular
the probability of direct thermal emission. In addition, we obtain as
a byproduct a description of the atomic and electronic structure of
the NaCl(100) surface, and of the nanocube corner vacancies with their
ionization and affinity levels, all of potential spectroscopic interest.

We did not attempt to simulate the process of electron attachment or
electron removal, since they are not relevant to our problem.  Actually,
calculations of electron attachment to a NaCl nanocluster already exist in
the literature based on quantum path-integral molecular dynamics methods,
using classical interatomic potentials, and by treating the quantum nature
of the extra electron in an effective way~\cite{landman85,barnett95}.

The paper is organized as follows.  In Sec.~\ref{sec:computational}
we will describe the computational method and the geometrical
aspects of the systems that are the subject of our calculations.
In Sec.~\ref{sec:results} we will present and discuss the electronic
structure for the neutral and charged NaCl(100) surface, and subsequently
that of the neutral and charged nanoclusters. In Sec.~\ref{sec:discussion}
we present the energetics of neutral atom emission processes, with an
estimation of thermal emission rates. Finally, Sec.~\ref{sec:conclusion}
will present our conclusions.

\section{Computational methods}\label{sec:computational}
We carried out ab initio electronic structure and total energy
calculations based on the density functional theory (DFT)~\cite{DFT}
in the generalized gradient approximation (GGA). We adopted
the plane wave pseudopotential approach~\cite{PWSCF}, using
norm-conserving pseudopotentials~\cite{MT} and a plane wave cutoff
of 40~Ry.  The Na pseudopotential was generated from the ionized
configuration ($3s^{0.6}\,3p^{0.1}\,3d^{0.1}$) and non-linear core
corrections~\cite{NLCC} were included.

Charged systems were simulated by adding one electron to the cluster
LUMO (conduction band in the infinite surface case) or subtracting one
electron from the cluster HOMO (valence band in the infinite surface case)
in presence of a uniform neutralizing background. Surfaces and isolated
nanocubes were simulated by periodically repeated slabs and nanocubes,
all replicas well separated by a sufficient amount of vacuum. All
calculations were spin polarized, and made use of the gradient-corrected
PW91 exchange-correlation functional~\cite{PW91}.

We started by computing the electronic structure, the equilibrium
lattice spacing (5.62~\AA) and the bulk modulus (22~GPa) of bulk
\emph{fcc} NaCl, in good agreement with experimental values (5.54~\AA\
and 28.6~GPa~\cite{haussul60}) and with previous calculations. The
calculated DFT energy gap is 5.42~eV, substantially smaller than the
experimental gap 8.97~eV (Ref.~\cite{roessler68}). We should point out
that this defect, which is standard in DFT calculations, does not cause
errors or problems with the accuracy of the total energy, so long as the
energetic ordering of levels is correct. Thus for example the total energy
calculation would develop problems if, say, the Kohn Sham $3s$ electronic
energy level of the neutral Na atom were to fall accidentally outside
the NaCl cluster energy gap, for that would wrongly make the neutral
atomic state unstable. We find no such occurrence in our calculations,
which can therefore be relied upon to a state of the art DFT accuracy.

As the next step, we calculated the electronic structure, total energy and
equilibrium atomic positions of neutral infinite NaCl(100) surface. For
that we used a slab consisting of 10 atomic layers, periodically
repeated and separated by 8~\AA\ of vacuum. Sampling of two dimensional
$k$-points was done by means of a Monkhorst-Pack mesh of 4$\times$4 in
the 2d Brillouin zone (BZ).  Surface electronic states were identified
based on their energy location inside the surface projected bulk energy
gap, and on their exponential decay from the surface towards the slab
center. The zero of Kohn-Sham electronic eigenvalues is taken to be the
vacuum level, initially extracted from the average value of the Hartree
potential in the geometric center of the vacuum region between the slabs,
after a careful extrapolation to the infinite distance limit.

Cubic NaCl nanocluster geometries were subsequently created by cutting out
cubelets from a perfect NaCl crystal, with an initial interatomic Na--Cl
distance of 2.81~\AA.  We studied neutral nanocubes of two sizes, a small
one Na$_4$Cl$_4$ made up of 8 atoms or 4 molecules (NC4), and a larger
one Na$_{32}$Cl$_{32}$ of 64 atoms or 32 molecules (NC32). Calculations
were carried in a cubic supercell of 10~\AA\ lateral side for the smaller
cube and 20~\AA\ for the larger one. We sampled the supercell BZ at the
$\Gamma$ point only. For the neutral nanocubes, we verified that the
residual interaction between periodic replicas is negligible. However,
for the charged cubes, the residual interaction is not negligible and
was removed by the technique described in Ref.~\cite{makov95}.

\section{Results}\label{sec:results}
We will begin this Section by describing the electronic states of
the infinite ideal neutral NaCl(100) surface, followed by results for
charged NaCl(100), where the extra electron and the extra hole are fully
delocalized.  Subsequently we will present results for the neutral and
charged nanocubes, and the calculated extraction energy curves.

\subsection{Neutral NaCl(100)}
Modeling the infinite flat NaCl(100) surface with a 10-layer slab, its
geometric structure was fully relaxed so as to minimize the calculated
DFT total energy. The largest relaxation occurred for the outer surface
atoms, the Na$^+$ surface ions relaxing inwards, the Cl$^-$ surface ions
outwards. As a result the electron charge distribution (carried almost
entirely by Cl$^-$ ions) spills out further into the vacuum.  Overall, the
zero temperature equilibrium neutral NaCl(100) geometry is thus predicted
to be slightly buckled, with a difference in height between Cl$^-$ surface
ions and Na$^+$ surface ions of 0.095~\AA. The computed surface energy is
13.17~meV/\AA$^2$ (211~erg/cm$^2$), in line with experimental estimates
and in striking agreement with Born-Mayer-Huggins-Fumi-Tosi classical
potential calculations, which yield about 200~erg/cm$^2$ at $T=$0.
(Ref.~\cite{zykova05}).

The electronic states of the slab were plotted in the 2d surface BZ, in
superposition with the surface projected bulk bands.  Possible NaCl(100)
surface electronic states were then identified by their energetic location
inside the surface projected bulk energy gap.

\begin{figure}\begin{center}
  \includegraphics[angle=270,width=0.8\columnwidth]{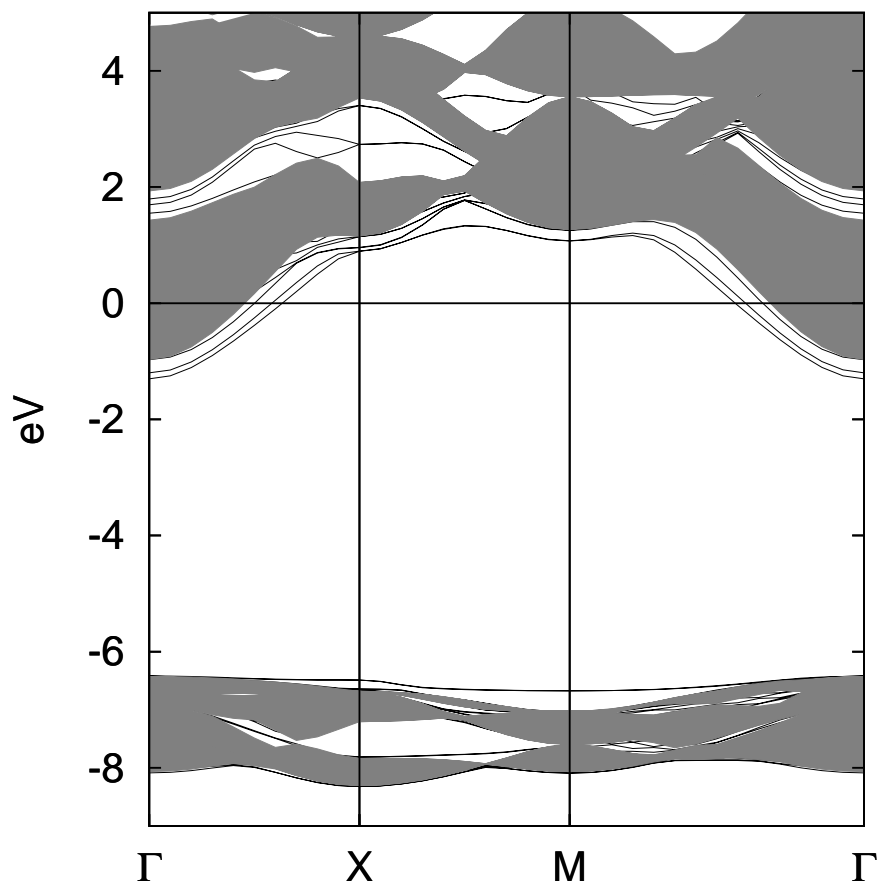}
  \caption{NaCl(100) surface band structure. The gray bands are the bulk
  projected band structure. The vacuum level is indicated by the horizontal
  line. The Cl $3s$ states (not shown) lie around $-$18~eV.}
  \label{fig:surfbands}
\end{center}\end{figure}

Our calculated surface electronic structure of the fully relaxed,
neutral NaCl(100) is shown in Fig.~\ref{fig:surfbands}. The edge of
the conduction band lies 1.02~eV below the vacuum level. Below that,
we find a pair of empty surface states, 1.35~eV and 1.46~eV below vacuum
respectively. The wavefunction of these states mainly consists of surface
Na $3s$ and $3p$ orbitals spilling out into vacuum -- strikingly unlike
the conduction states of bulk NaCl, which have instead a sizable Cl
$3p$ character.  Qualitatively, these neutral surface states suggests
electron affinity levels of NaCl(100) of a very similar nature, namely
surface Na $3s$ and $3p$. Their significance however is obvously
not quantitative, because these states are not occupied by an actual
electron. The experimentally reported electron affinities of NaCl in
fact range from 0.5~eV (Refs.~\cite{mott50}  and~\cite{poole75}) to 1~eV
(Ref.~\cite{wright48}), roughly 35\% to 70\% of the calculated empty
surface state energy value below vacuum zero.

By charging the slab with an extra electron, we attempted to compute
the electron affinity as the total energy difference between the slab
with an extra electron and the neutral slab $A = E(N+1) - E(N)$. This
should work in principle so long as the sequence of the electronic
levels is preserved. However we found that within DFT-PW91, the extra
electron level rises above the conduction band edge. As a results, the
SCF procedure does not converge.  This we believe is an effect of the
incomplete cancellation of self-interaction in DFT, and thus an artifact,
reflecting the well known difficulty that approximate XC functionals
have in binding extra electrons~\cite{shore77,perdew81}.  Due to that
we do not have a reliable calculation of the surface electron affinity
levels of NaCl(100), and all we can state is that they should lie between
the empty surface state levels at  1.35~eV and 1.46~eV below vacuum,
and vacuum zero.

Coming to the filled states, the surface projected valence band edge
lies $\sim$6.4~eV below vacuum. Due to the DFT underestimation of the
energy gap, the true valence band edge actually lies $\sim$9$\div$10~eV
below the vacuum level, as suggested by the experimental work
function~\cite{wright48}. Just above the calculated valence band edge
we find for neutral NaCl(100) a filled surface state, whose existence
is restricted to the neighborhood of the M point in the 2d BZ.  Like the
valence band states, this surface state has a Cl $3p$ character. We
repeated the calculation by removing one electron and we found that the
hole state is delocalized over all surface Cl $3p_z$ orbitals.

\subsection{Neutral NaCl nanocubes}
It is known from experiments~\cite{martin83,conover88} that gas phase
ionized NaCl clusters exhibit in equilibrium ``magic sizes'' corresponding
to small cubelets exposing (100) facets, whose general formula is
[Na$_n$Cl$_{n-1}$]$^{+}$ or [Na$_n$Cl$_{n+1}$]$^{-}$.  Even if neutral
clusters cannot be directly detected in mass-spectroscopy experiments
because of the destabilizing nature of charging, they are nonetheless
predicted~\cite{conover88} with cubic shape, and a different general
formula Na$_{2n}$Cl$_{2n}$. Equilibrium charged clusters (which are
observed) possess an odd number of ions on the cubic edges, as opposed to
neutral clusters which have an even number of ions. Positively charged
clusters, in this idealized picture, have only Na$^+$ corner ions,
no chlorine corners at all. Conversely, negatively charged clusters
have Cl$^-$ corner ions only, and no sodium corners at all. Neutral
Na$_{2n}$Cl$_{2n}$ clusters instead have four Na$^+$ and four Cl$^-$
corner ions. Our aim here is to predict electronic properties and
energies valid for NaCl clusters up to macroscopic sizes, which are
not destabilized by a single electron or a single hole. Thus we will
restrict our calculations to Na$_{2n}$Cl$_{2n}$ nanocubes, even in the
charged case.

In order to get a first account of the energetics associated with the
structural relaxation of the nanocubes, we carried out the structural
relaxation of the two nanocubes Na$_4$Cl$_4$ and Na$_{32}$Cl$_{32}$,
starting from the perfect cubic structures, generated according to the
prescription of Sec.~\ref{sec:computational}.  The deviations of atomic
coordinates from idealized bulk-like positions are largest for the corner
atoms. The Na corner ions relax strongly inwards, towards the center of
the cube, while the Cl corner ions relax weakly outwards.  The corner
atom displacements are shown in Tab.~\ref{tab:relax}, together with the
energy difference between the relaxed structures and the ideal cubes.
The pattern of displacements of the Na$^+$ and Cl$^-$ corner ions follows
the same trends as for the NaCl(100) relaxation, namely Na$^+$ corner
atoms move toward higher electronic density regions, while the Cl$^-$
corner atoms relax away from it.

\begin{table}\begin{center}
  \begin{tabular}{ccccc}
  \hline\hline
  & Small cube & \hspace{0.5cm} & Large cube & \\
  & (Na$_4$Cl$_4$) && (Na$_{32}$Cl$_{32}$) & \\
  \hline
  Na corner displ. & 0.35~\AA && 0.54~\AA & (inward) \\
  Cl corner displ. & 0.02~\AA && 0.21~\AA & (outward) \\
  $\Delta E$       & $-$0.65~eV && $-$1.59~eV & \\
  $\Delta\mathcal{E}_\mathrm{LUMO}$ & +0.15~eV && +0.10~eV & ($\times$~1)\\
  $\Delta\mathcal{E}_\mathrm{HOMO}$ & $-$0.71~eV && $-$0.45~eV&($\times$~3)\\
  \hline\hline
  \end{tabular}
  \caption{Structural relaxation of the neutral nanoclusters, starting
  from the ideal cubic structures. The final structure has $\mathrm{T}_d$
  symmetry. $\Delta E$ is the energy difference between the relaxed
  structures and the starting one. $\Delta\mathcal{E}_\mathrm{LUMO}$ and
  $\Delta\mathcal{E}_\mathrm{HOMO}$ are the energy difference of the HOMO
  and LUMO with respect to the initial ideal structure with bulk NaCl
  distances. The degeneracy of LUMO and HOMO nanocube levels is given in
  the last column.}
  \label{tab:relax}
\end{center}\end{table}

The relaxation process lowers the symmetry of the nanocubes from
$\mathrm{O}_h$ to $\mathrm{T}_d$, the tetrahedral group.  The HOMO is
a three-fold degenerate linear combination of corner Cl $3p$ orbitals,
their lobes pointing along the diagonals of the cube ($[111]$ directions).
The LUMO is non-degenerate and consist of a linear combination of Na
corner atoms $3s$ and $3p$ orbitals. The LUMO wavefunction is, as could
be expected, very much more diffuse outside the nanocluster than the HOMO.

The neutral nanocube HOMO and LUMO states might be hoped to provide,
as the surface states of the previous Section did, a qualitative
indication of the ionization and affinity levels of the nanocube,
after an electron is removed or added to the nanocube. However, after
charging the nanocube, we will find that in this case levels actually
fail to provide a good reference state for computing the extraction
energies. The reason is twofold.  First, removal of an electron from the
HOMO is accompanied by a Jahn-Teller (JT) distortion and the symmetry of
the nanocube is lowered. The energy gain associated with the JT distortion
can be large in the smallest nanocubes. A precise calculation of this
artificial JT energy gain within DFT is beyond the scope of this article.
On the contrary, addition of an electron to the LUMO is less problematic
since on account of the lack of true degeneracy it only gives rise to a
relaxation, or pseudo JT effect, where the energy gain is smaller with
respect to the JT case.  Second, these highly symmetrical HOMO and LUMO
states, perfectly delocalized over opposite corners as they are because
of symmetry, are unrealistic.  In fact as soon as one corner atom is
even slightly displaced outwards, the extended nature is removed and
the extra electron (hole) wave function collapses to become strongly
localized precisely on that atom.  Based on these considerations, the
effective initial state energy is calculated by assuming a very small
outward displacement which breaks cubic symmetry in the initial state too,
and extrapolating to zero the initial displacement.

\subsection{Electron addition: corner neutral Na atom detachment}
We simulated the detachment of a corner neutral Na atom by donating an
electron to the small cube NC4$^-$ and to the large one NC32$^-$. As
discussed in the previous sections, the Na$^+$ corner ions constitute
the affinity site for the extra electron, whose wavefunction collapses
onto a single Na corner ion as this is even slightly pulled outwards. As
a result, already at the earliest stage of detachment the Na$^+$ corner
ion becomes naturally neutralized (reduced) to its neutral state.

\begin{table}\begin{center}
  \begin{tabular}{lcccc}
  \hline\hline
  & small cube & \hspace{0.5cm} & large cube \\
  \hline
  Na extraction $\Delta E_\mathrm{Na}\hspace{0.5cm} $ & 0.62~eV && 0.68~eV \\
  (electron addition) & && & \\
  Cl extraction $\Delta E_\mathrm{Cl}$ & 0.80~eV && 0.78~eV \\
  (electron removal) & && & \\
  \hline\hline
  \end{tabular}
  \caption{Energy necessary to extract a neutral atom from the corner of
  a nanocube.}
  \label{tab:energies}
\end{center}\end{table}

\begin{figure}\begin{center}
  \includegraphics[width=0.8\columnwidth]{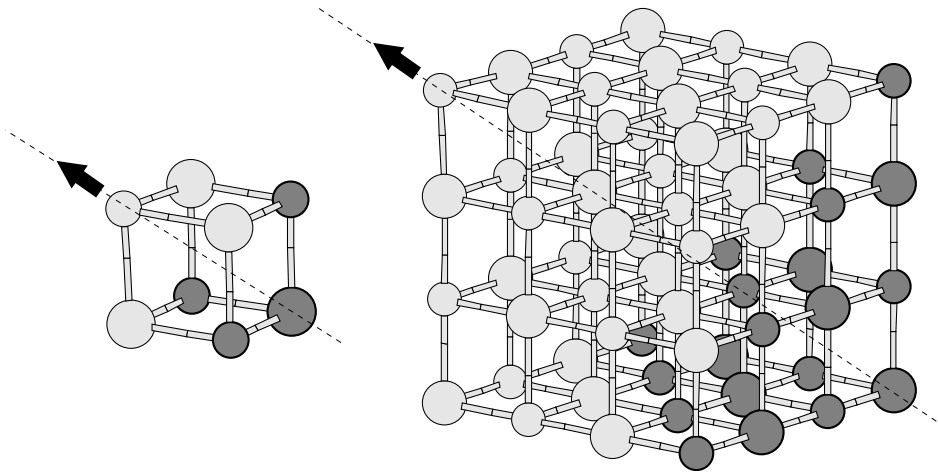}
  \caption{Left: small cube $[$Na$_4$Cl$_4]^-$. Right: large cube
  $[$Na$_{32}$Cl$_{32}]^-$. Small circles: sodium; Large circles:
  chlorine. Dark circles: fixed atoms (both Na and Cl). The arrow
  indicates the direction of extraction of the Na corner atom.}
  \label{fig:nanocubi}
\end{center}\end{figure}

To calculate the energy cost of detachment we gradually displaced the
Na atom outwards in the $[111]$ direction away from the corner of the
nanocube.  The displacement was measured from the ideal corner coordinates
of the perfect cube, taken as the reference geometry. Relaxation of
the other ions was allowed during detachment. Actually, in order to
simulate the mechanical rigidity of a much bigger cube or corner, only
ions closest to the resulting corner vacancy were allowed to relax,
whereas the remaining ions (roughly 1/3 of the total) were kept fixed
as indicated by the different shading in Fig.~\ref{fig:nanocubi}. We
relaxed the structure adiabatically after each outward displacement of
the corner atom, increased in steps from 0.1~\AA\ up to 5.0~\AA.

\begin{figure}\begin{center}
  \includegraphics[angle=270,width=0.9\columnwidth]{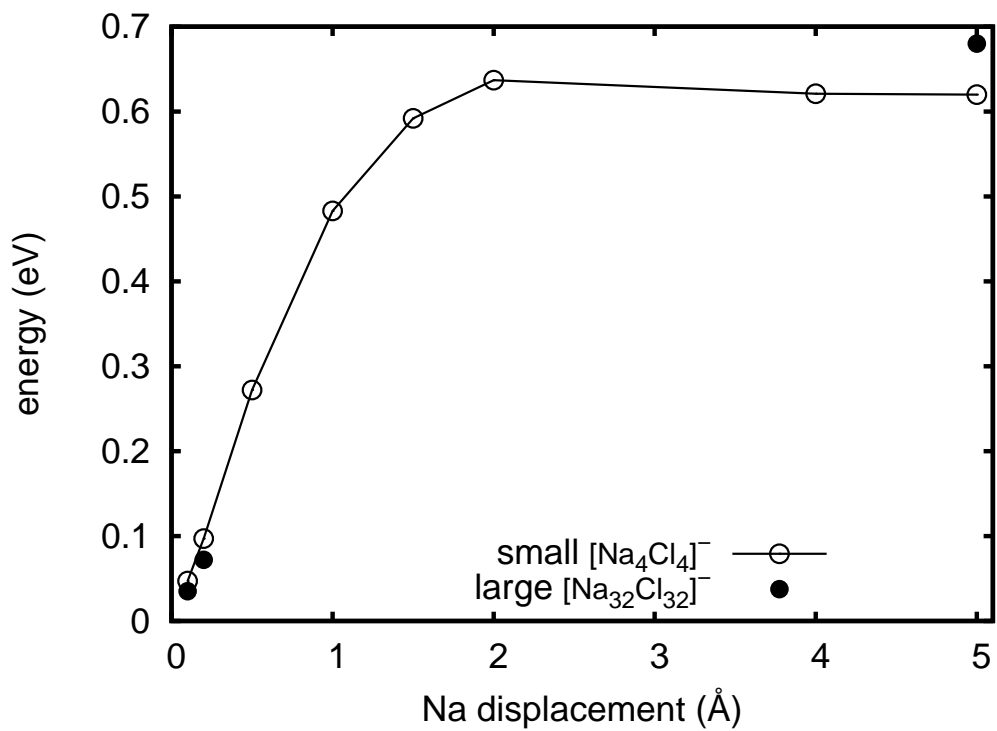}
  \caption{Na extraction energy profile (electron addition). The total
  energy of the relaxed nanocubes is plotted against the outwards
  displacement of one Na corner atom along the $[111]$ direction.}
  \label{fig:na_extraction}
\end{center}\end{figure}

The total energy change is followed by relaxing adiabatically the
remaining atomic positions so as to minimize energy as a function of the
corner Na nucleus extraction coordinate. Extraction is complete when
the total energy levels off at some large pull-off distance which we
take to be 5~\AA. The total energy is found to increase monotonically,
leveling off for an outward displacement of $\sim$4~\AA\ and higher
(Fig.~\ref{fig:na_extraction}). The extraction energy is evaluated
as a difference between the total energy for a displacement of 5~\AA\
and the total energy for zero displacement. The results are summarized
in Tab.~\ref{tab:energies}. The energy required to extract a neutral
Na atom, by addition of an electron is calculated to be $\sim$0.6~eV,
in fairly good (even if in our view somewhat fortuitous, given the
approximations implied by the empirical estimates) agreement with the
previous estimate of 0.4~eV (Refs.~\cite{martin83,diefenbach85,galli86}).

\begin{figure}\begin{center}
  \includegraphics[width=0.6\columnwidth]{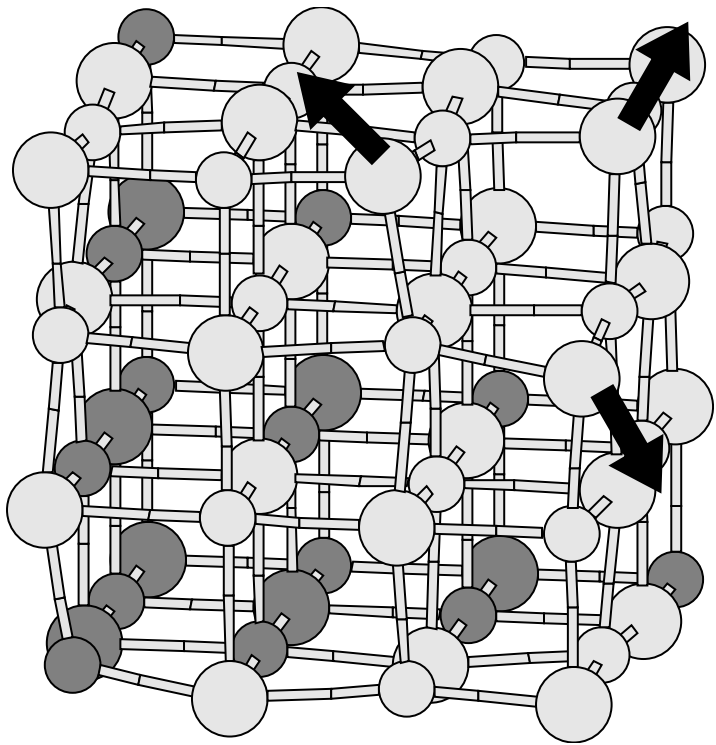}
  \caption{Relaxation pattern around the Na corner vacancy (top right) in
  the large cube when the neutral Na atom is fully extracted
  ($[$Na$_{31}$Cl$_{32}$]$^-$). Small circles: sodium; large circles:
  chlorine. Dark circles: fixed atoms (both Na and Cl).}
  \label{fig:na_relaxation}
\end{center}\end{figure}

The detachment energetics we just presented is in principle affected by
nanocube finite-size effects. However, the near coincidence of results
of the small and large cubes indicates that finite-size errors arising
from various sources are relatively unimportant. This also agrees with
the localized nature of distortions and of electronic states involved.
Because of that we did not attempt to compute larger nanocubes, due to the
much larger computational costs, but also comforted by the convergence
observed. Our small size results are we believe fully representative of
the physics of even a macroscopic cubic corner.

More subtly, the incomplete cancellation of electron self-interaction
in DFT might constitute another source of error. A different degree of
localization of the extra electron in the initial and final state will
introduce an error, due to a correspondingly different self-interaction of
the two electron states: the initial one spread on the whole cluster and
the final one in the neutral Na atom only. However, as explained earlier,
we generate our effective initial state by introducing a very small
displacement which breaks the ideal cubic symmetry in the initial state
too. With that, the localization of the electronic state is no longer
very different in the initial and final states, implying that also the
self-interaction error is very much reduced. The effect is in any case
not a large one to start with. For an appreciation of the magnitude,
total energy of the perfectly cubic NC32$^-$ is $E_0=-$1687.173~eV, that
of the same nanocube after a slight corner Na extraction (extrapolated
to zero) is $E_0'=-$1687.211~eV, and the final one after full Na atom
extraction $E_5'=-$1686.531~eV. The difference $E_0 - E_0'=$~38~meV
(which also contains the self-interaction), is negligible compared  the
Na extraction energy $\Delta = E_5'-E_0' =$~0.68~eV.

\begin{figure}\begin{center}
  \includegraphics[angle=270,width=0.8\columnwidth]{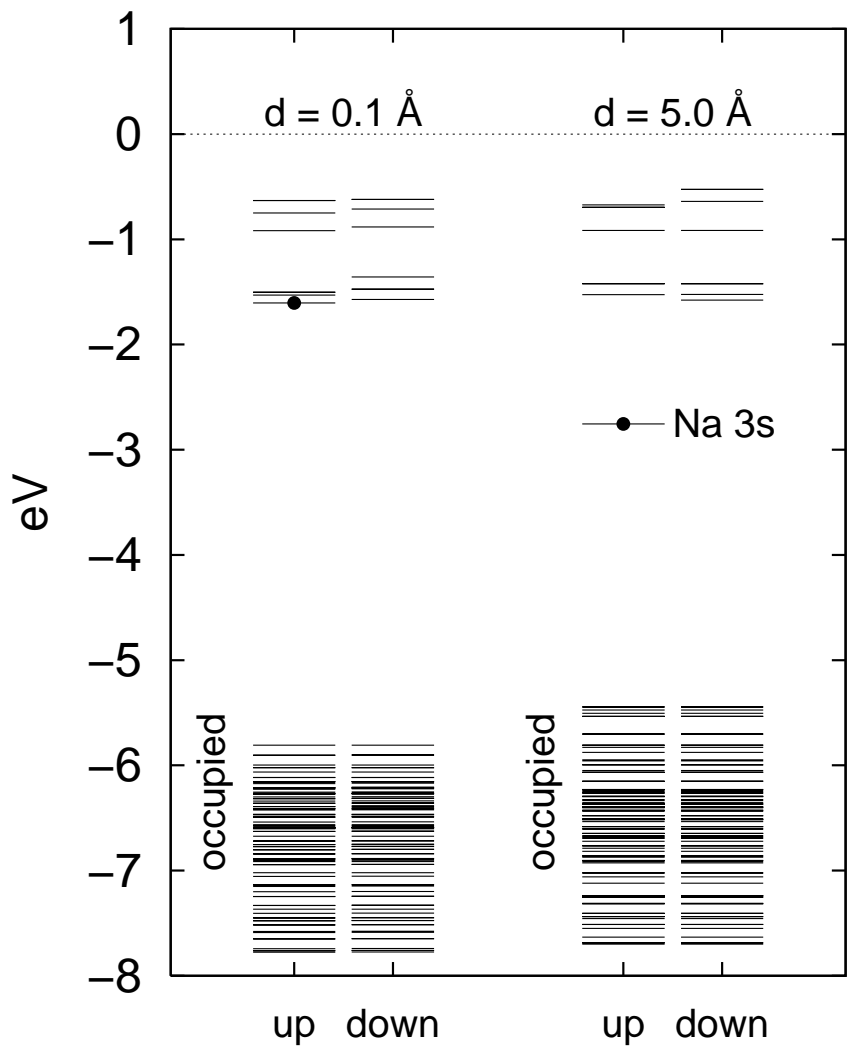}
  \caption{Na extraction (electron addition): energy levels (in eV)
  from the vacuum level (set to zero) for the large $[$Na$_{32}$Cl$_{32}]^-$
  nanocube. Left column: corner Na displacement 0.1~\AA; right column:
  corner Na displacement 5.0~\AA\ (fully extracted). The levels below
  $-$5.5~eV are fully occupied. The solid circle indicates the level of
  the extra electron.}
  \label{fig:ele_levels}
\end{center}\end{figure}

Finally, we examined the atomic relaxation pattern and the electronic
structure of the Na corner vacancy, after the neutral Na atom is fully
extracted. All atomic relaxations (Fig.~\ref{fig:na_relaxation} are mostly
localized in vicinity of the vacancy. The three first neighbors Cl$^-$
ions relax outwards in such a way as to minimize their electrostatic
repulsion. We show in Fig.~\ref{fig:ele_levels}, the energy levels of
the large cube $[$Na$_{32}$Cl$_{32}]^-$) with an extra electron, in the
initial and final state ($[$Na$_{31}$Cl$_{32}]^-$ + Na).  Initially, the
electron is trapped in a corner state $\sim$1.6~eV below vacuum. After
Na extraction, the extra electron sits on the Na $3s$ atomic level which
lies $\sim$2.75~eV below vacuum. We note that this level is too high in
energy, for in reality the valence electron energy of a neutral Na atom in
vacuum below vacuum zero should be about equal the Na ionization potential
(5.1~eV).  As is well known, this is a standard DFT problem. However,
as far as total energy differences are concerned, DFT is generally
accurate and reliable, at least so long as the self consistent level
configuration has the correct ordering, which it does in our case.

In the final state, negative nanocube plus detached neutral Na atom,
there is also a second group of levels lying in the gap, around $-$5.5~eV
-- that is about 0.3~eV above the bulk valence states.  These are Cl
based defect states associated with the relaxation the three Cl ions
neighboring the Na corner vacancy.

\subsection{Electron removal: corner Cl detachment}
We simulated the detachment of a corner Cl atom by removing an electron
from the nanocube, generating both a small positive cluster NC4$^+$ and
a large one  NC32$^+$. The overall mechanics is similar to the previous
section.  In the case of Cl, an electron is naturally removed from a
Cl$^-$ corner atom, neutralizing (oxidizing) it to its neutral state.

\begin{figure}\begin{center}
  \includegraphics[angle=270,width=0.9\columnwidth]{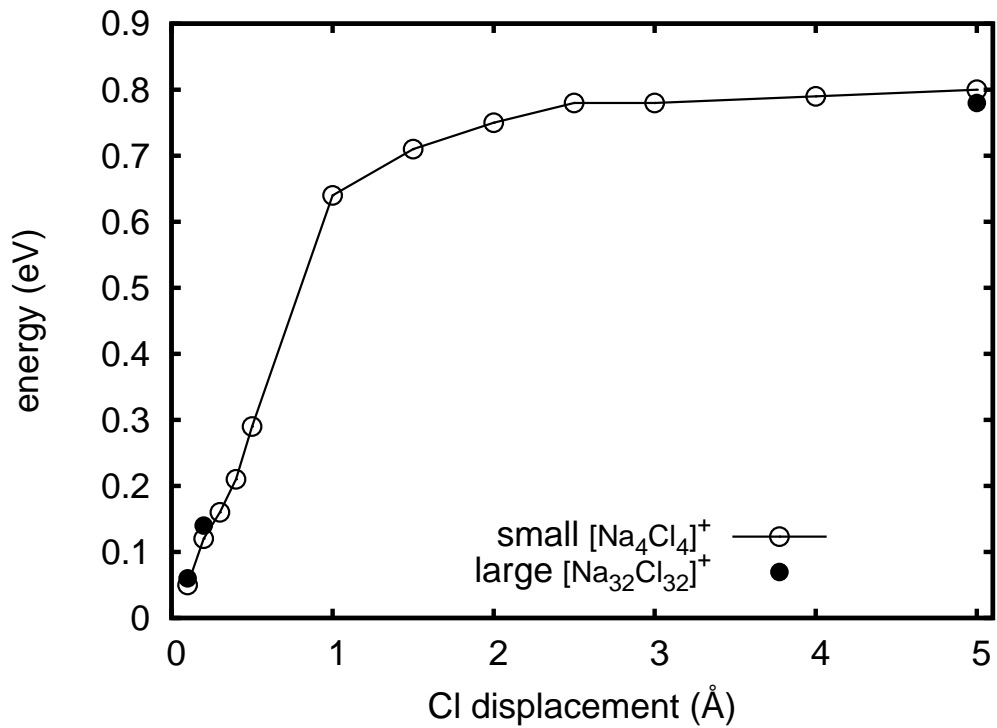}
  \caption{Cl extraction energy (electron removal).}
  \label{fig:cl_extraction}
\end{center}\end{figure}

We followed a procedure specular to that described in the previous
section, now for the Cl corner atom. Similarly to the case of Na, the
total energy upon Cl detachment was found to increase monotonically
as a function of the Cl outward displacement along $[111]$,
leveling off for an outward displacement of $\sim$4~\AA\ and higher
(Fig.~\ref{fig:cl_extraction}). The extraction energy is evaluated as
a difference between the total energy for a displacement of 5~\AA\ and
the total energy for zero displacement.  Like in the previous case, we
considered the sources of errors, and found them to be equally tolerable.

Our resulting Cl extraction energy (Tab.~\ref{tab:energies})
$\Delta E_\mathrm{Cl}\sim$0.8~eV however is in contrast
with the earlier semiempirical estimate of 0.04~eV, 20 times
smaller~\cite{martin83,diefenbach85}.  That estimate however had been
obtained attributing the binding energy of a neutral Cl atom entirely to
dispersion forces, much smaller in Cl than in Na because their atomic
polarizabilities differ roughly by a factor 3.  Our result shows on
the contrary that the energy cost to extract a neutral Cl or a neutral
Na atom are close. This suggests that the main contribution could be
the formation energy of the corner vacancy.  This energy cost could be
reduced if the Cl atom is adsorbed on top of another Cl atom, as shown
recently in Ref.~\cite{li06}.

\begin{figure}\begin{center}
  \includegraphics[angle=270,width=0.8\columnwidth]{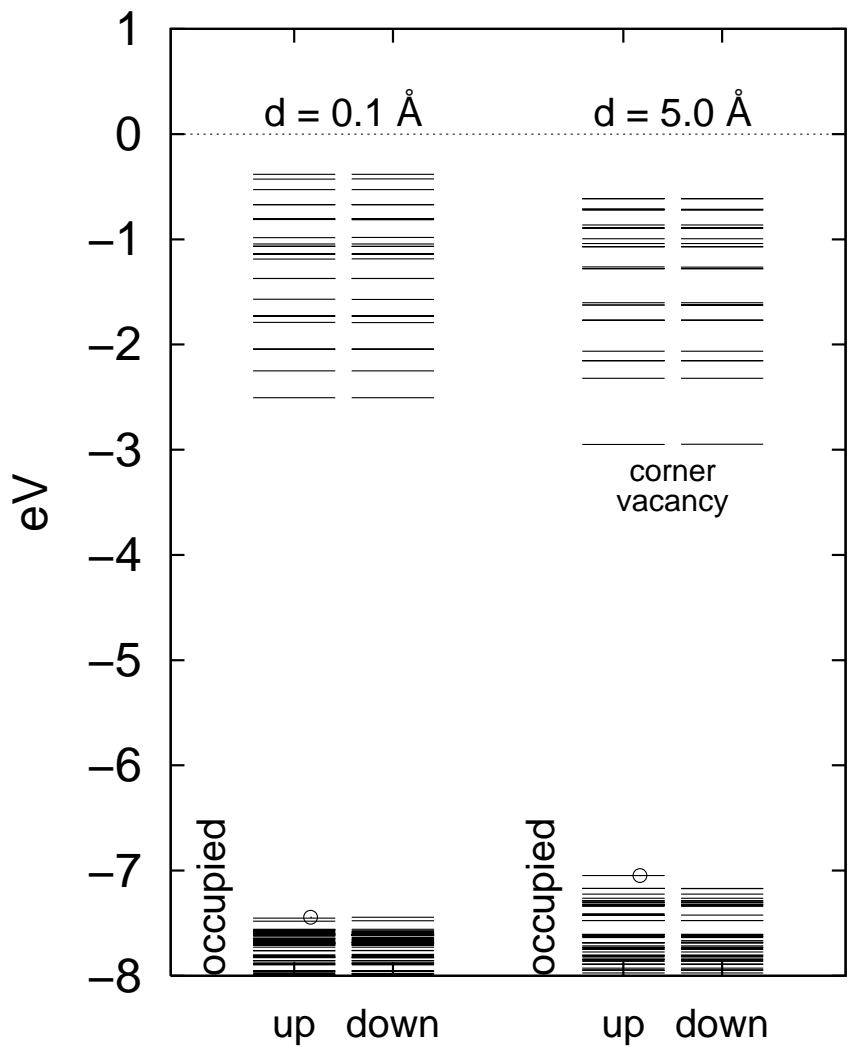}
  \caption{Cl extraction (electron removal): energy levels (in eV) from
  the vacuum level (set to zero) for the large ($[$Na$_{32}$Cl$_{32}]^+$
  nanocube. Left column: corner Cl displacement 0.1~\AA; right column:
  corner Cl displacement 5.0~\AA\ (fully extracted). The open circle
  indicates the position of the hole, otherwise the levels below $-$7~eV
  are fully occupied.}
  \label{fig:hole_levels}
\end{center}\end{figure}

As in the case of the Na$^+$ vacancy, atomic relaxations (not shown)
are now mostly localized in the vicinity of the Cl$^-$ vacancy. The
first three neighboring Na$^+$ ions relax inwards.  We show in
Fig.~\ref{fig:hole_levels}, the energy levels of the large cube
($[$Na$_{32}$Cl$_{32}]^+$) with an extra hole, in the initial and final
state.  Initially, the hole is trapped in a corner state $\sim$7.5~eV
below vacuum.  In the final state, the hole sits in one of the Cl $3p$
atomic levels, $\sim$7~eV below vacuum. After the Cl$^-$ is extracted, a
corner vacancy empty electronic state $\sim$3~eV appears below the vacuum
level.  This level might act as a final state for optical transitions
from the neighboring  Cl$^-$  ions, similar to optical transitions in
bulk or surface F--centers. Experimentally a surface F--center empty
level is known to lie $\sim$2~eV below vacuum.~\cite{zielasek00}

\section{Discussion}\label{sec:discussion}
We calculated the detachment energetics of detachment of neutral atoms
from NaCl nanocube corners after charging the nanocube with an electron or
with a hole. The energies of detachment -- $\sim$0.6~eV for neutral Na,
$\sim$0.8~eV for neutral Cl -- are similar, and altogether relatively
small.

From a practical point of view, among the two processes, the electron
induced extraction of Na atoms from NaCl nanocubes is more important for
two reasons.  First, an extra electron coming from vacuum can gain energy
by spontaneously attaching to the cluster and occupying the empty surface
states which give rise to a positive electron affinity of NaCl(100),
of order $\sim$1~eV, to be compensated by an energy cost of 0.6~eV to
extract a Na corner atom. By contrast, electron removal requires an
energy as large as the work function ($\sim$9~eV).  Second, detachment
of neutral Na atoms from charged NaCl nanocubes could provide a cheap
abundant source of very reactive atoms. After detachment, a neutral
alkali atom could react in gas phase with a water molecule yielding
sodium hydroxide and molecular hydrogen:
\begin{displaymath}
 \mathrm{Na}(g) + \mathrm{H}_2\mathrm{O}(g) \rightarrow
 \mathrm{NaOH}(s) + (1/2)\mathrm{H}_2(g)
\end{displaymath}
a reaction which is very exothermic, with an enthalpy of reaction of
$-$247~kJ/mol or ($-$2.56~eV per Na atom).~\cite{CRC}  Combined with
our calculated Na extraction energy, the net balance is still exotermic,
$\sim$190~kJ/mol ($\sim$2~eV per Na atom).

Even assuming the starting neutral NaCl nanoclusters, and an abundant
"beam" of (nearly) zero-energy electrons to be given, the main limiting
factors in this scheme are still two. They are the uncertain yield of
electron attachment, and the small escape rate of Na atoms.  Electron
attachment depends crucially on how the electron is donated to the cluster
and on the microscopic mechanisms of excess energy disposal, a subject far
beyond the scope of this paper. The Na atom thermal escape rate (average
time between single atom emission events) can be estimated easily as
\begin{displaymath}
  \tau = \Omega^{-1} \exp\left[\Delta E_\mathrm{Na}/k_B T\right]
\end{displaymath}
Assuming an attempt frequency $\Omega$ of $\sim$1~THz, the neutral Na
atom escape rate is very small, of the order of one atom every 1.8~$s$
at room temperature.  Of course, that could be substantially increased
by heating.

\section{Conclusion}\label{sec:conclusion}
We calculated the extraction process of a neutral Na (Cl) atom
from NaCl nanoclusters to which one electron has been donated
(removed).  For the neutral Na atoms, a relatively low extraction
energy of the order of 0.6~eV, about 50 \% higher than previously
suggested~\cite{martin83,diefenbach85,galli86} is brought out
by our calculation. The extraction energy of a neutral Cl atom
is of the order of 0.8~eV, about 20 times larger than earlier
suggestions~\cite{martin83,diefenbach85}.  It is suggested that neutral
Na atoms that will to some extent be thermally emitted by corner sites
of negatively charged NaCl nanocubes could provide a cheap source of
very reactive halogen atoms.

\ack
This project is sponsored by Italian Ministry of University and Research,
through PRIN-2006022847 and by INFM, through ``Iniziativa Trasversare
Calcolo Parallelo''. Calculations were performed on the SP5 and CLX
clusters at CINECA, Casalecchio (Bologna). We acknowledge suggestions by,
and discussions with, F. Stellacci, who also pointed out the potential
importance of cheap reactive atom sources.




\section*{References}
\begin{thebibliography}{90}
\bibitem{CRC}
   Lide D R
   1994 \emph{CRC Handbook of Chemistry and Physics, 74th ed.}
   (CRC Press, London)

\bibitem{zielasek00}
   Zielasek V, Hildebrandt T and Henzler M
   2000 \emph{Phys. Rev. B} \textbf{62} 2912

\bibitem{georgiew88}
   Georgiev M
   1988 \emph{F--centers in alkali halides} (Springer, Berlin)

\bibitem{li92}
   Li X, Beck R D and Whetten R L
   1992 \emph{Phys. Rev. Lett.} \textbf{68} 3420

\bibitem{puchin93}
   Puchin V E, Shluger A L and Itoh N
   1993 \emph{Phys. Rev. B} \textbf{47} 10760

\bibitem{puchin94}
   Puchin V E, Shluger A L, Nakai Y and Itoh N
   1994 \emph{Phys. Rev. B} \textbf{49} 11364

\bibitem{rohlfing05} 
  Rohlfing M, Wang N P, Kr\"uger B and Pollmann J
  2005 \emph{Surf. Sci.} \textbf{593} 19

\bibitem{meyer05}
   Repp J, Meyer G, Paavilainen S, Olsson F E and Persson M,
   2005 \emph{Phys. Rev. Lett.} \textbf{95} 225503

\bibitem{conover88}
   Conover C W S, Yang Y A and Bloomfield L A
   1988 \emph{Phys. Rev. B} \textbf{38} 3517

\bibitem{martin83}
   Martin T P,
   1983 \emph{Phys. Repts.} \textbf{95} 168

\bibitem{diefenbach85}
   Diefenbach J and Martin T P
   1985 \emph{J. Chem. Phys.} \textbf{83} 4585

\bibitem{galli86}
   Galli G, Andreoni W and Tosi M P
   1986 \emph{Phys. Rev. A} \textbf{34} 3580

\bibitem{landman85}
   Landman U, Scharf D and Jortner J
   1985 \emph{Phys. Rev. Lett.} \textbf{54} 1860

\bibitem{barnett95}
   Barnett R N, Cheng H-P, Hakkinen H and Landman U
   1995 \emph{J. Phys. Chem.} \textbf{99} 7731

\bibitem{DFT}
   Hohenberg P and Kohn W, 1964 \emph{Phys. Rev.} \textbf{136} B864;
   Kohn W and Sham L J, \emph{Phys. Rev.} \textbf{140} A1133

\bibitem{PWSCF}
   Baroni S, Dal Corso A, de Gironcoli S, Giannozzi P, Cavazzoni C,
   Ballabio G, Scandolo S, Chiarotti G, Focher P, Pasquarello A,
   Laasonen K, Trave A, Car R, Marzari N and Kokalj A
   \textsl{http://www.pwscf.org}

\bibitem{MT}
   Troullier N and Martins J L
   1991 \emph{Phys. Rev. B} \textbf{43} 1993

\bibitem{NLCC}
   Louie S G, Froyen S and Cohen M L
   1982 \emph{Phys. Rev. B} \textbf{26} 1738

\bibitem{PW91}
   Wang Y and Perdew J P,
   1991 \emph{Phys. Rev. B} \textbf{43} 8911;
   Perdew J P and Wang Y,
   1992 \emph{Phys. Rev. B} \textbf{45} 13244

\bibitem{haussul60}
  Hauss\"ul S
  1960 \emph{Z. Phys.} \textbf{159} 223

\bibitem{roessler68}
   Roessler D M and Walker W C
   1968 \emph{Phys. Rev.} \textbf{166} 599

\bibitem{makov95}
   Makov G and Payne M C
   1995 \emph{Phys. Rev. B} \textbf{51} 4014

\bibitem{zykova05}
   Zykova-Timan T, Ceresoli D, Tartaglino U and Tosatti E
   2005 \emph{J. Chem. Phys.} \textbf{123} 164701

\bibitem{mott50}
   Mott N F and Gurney R W
   1950 \emph{Electronic Processes in Ionic Crystals, 2nd ed.} (Oxford)

\bibitem{poole75}
   Poole R T, Liesegang J, Leckey R C G and Jenkin J G
   1975 \emph{Phys. Rev. B} \textbf{11} 5190

\bibitem{wright48}
   Wright D A
   1948 \emph{Proc. Phys. Soc.} \textbf{60} 13

\bibitem{shore77}
  Shore H B, Rose J H and Zaremba E
  1977 \emph{Phys. Rev. B} \textbf{15} 2858

\bibitem{perdew81}
  Perdew J P and Zunger A
  1981 \emph{Phys. Rev. B} \textbf{23} 5048

\bibitem{li06}
  Li B, Michaelides A, Scheffler M,
  2006 Phys. Rev. Lett. \textbf{97} 046802


\end{thebibliography}
\end{document}